\documentclass{article}
\usepackage{spconf,amsmath,graphicx}
\usepackage{amsfonts}
\usepackage{float}
\usepackage{booktabs}

\title{A Generative Model for Hallucinating Diverse Versions of\\ Super Resolution Images}
\name{Mohamed Abid, Ihsen Hedhli, Christian Gagné$^*$ }
\address{IID, Université Laval, $^*$Canada-CIFAR AI Chair, Mila}
\begin{document}
%
\maketitle
\begin{abstract}
Traditionally, the main focus of image super-resolution techniques is on recovering the most likely high-quality images from low-quality images, using a one-to-one low- to high-resolution mapping. Proceeding that way, we ignore the fact that there are generally many valid versions of high-resolution images that map to a given low-resolution image. We are tackling in this work the problem of obtaining different high-resolution versions from the same low-resolution image using Generative Adversarial Models. Our learning approach makes use of high frequencies available in the training high-resolution images for preserving and exploring in an unsupervised manner the structural information available within these images. Experimental results on the CelebA dataset confirm the effectiveness of the proposed method, which allows the generation of both realistic and diverse high-resolution images from low-resolution images.
\end{abstract}
\begin{keywords}
Super-resolution, Image Hallucination, Generative Adversarial Models.
\end{keywords}
\section{Introduction}
\label{sec:intro}

Image super-resolution (SR) is an ill-posed problem, since each low-resolution (LR) image can have a practically infinite number of corresponding images in the high-resolution (HR) domain. However, traditional SR techniques rely on a one-to-one schema, aiming at generating (or reconstructing) only one HR image from a LR image, generally the most likely one. However, they still do not account for all other possible mappings. In practice, in applications such as microscopy or medical imaging, where experts rely on SR methods, having access to more than one possible solution can lead to better strategies where different outcomes, and even the certainty on these outcomes, are considered.

In this paper, we seek to perform a one-to-many schema by generating a diverse set of HR images out of one single LR image. For that, we are proposing to use a Generative Adversarial Network (GAN) that recovers an HR image from a very LR version of the same image, meanwhile being able to hallucinate a wide variety of other acceptable HR versions of this image. In the recent literature, some exploratory SR has been investigated, for example relying on SR methods that are either controlled by a user interface to manually manipulate the variance and periodicity of textures \cite{bahat}, or by using semantically guided style imposition \cite{deepsee}. The diversity generated in these methods are supervised, by either a human or a specific criterion, which is in contrast to our method where hallucination is conducted in an unsupervised way. Another approach such as \cite{pulse} searches in a pretrained GAN’s manifold for a set of high-resolution images that maps, when downscaled, to the same low-resolution image. While this method yields great quality samples, it is an iterative procedure that requires a significant amount of time to generate each image. Meanwhile our method is a one pass-prediction which works for general images and is not dependent on a pretrained GAN (e.g., StyleGAN). 
\begin{figure}[t]
    \centering
    \includegraphics[width=\linewidth]{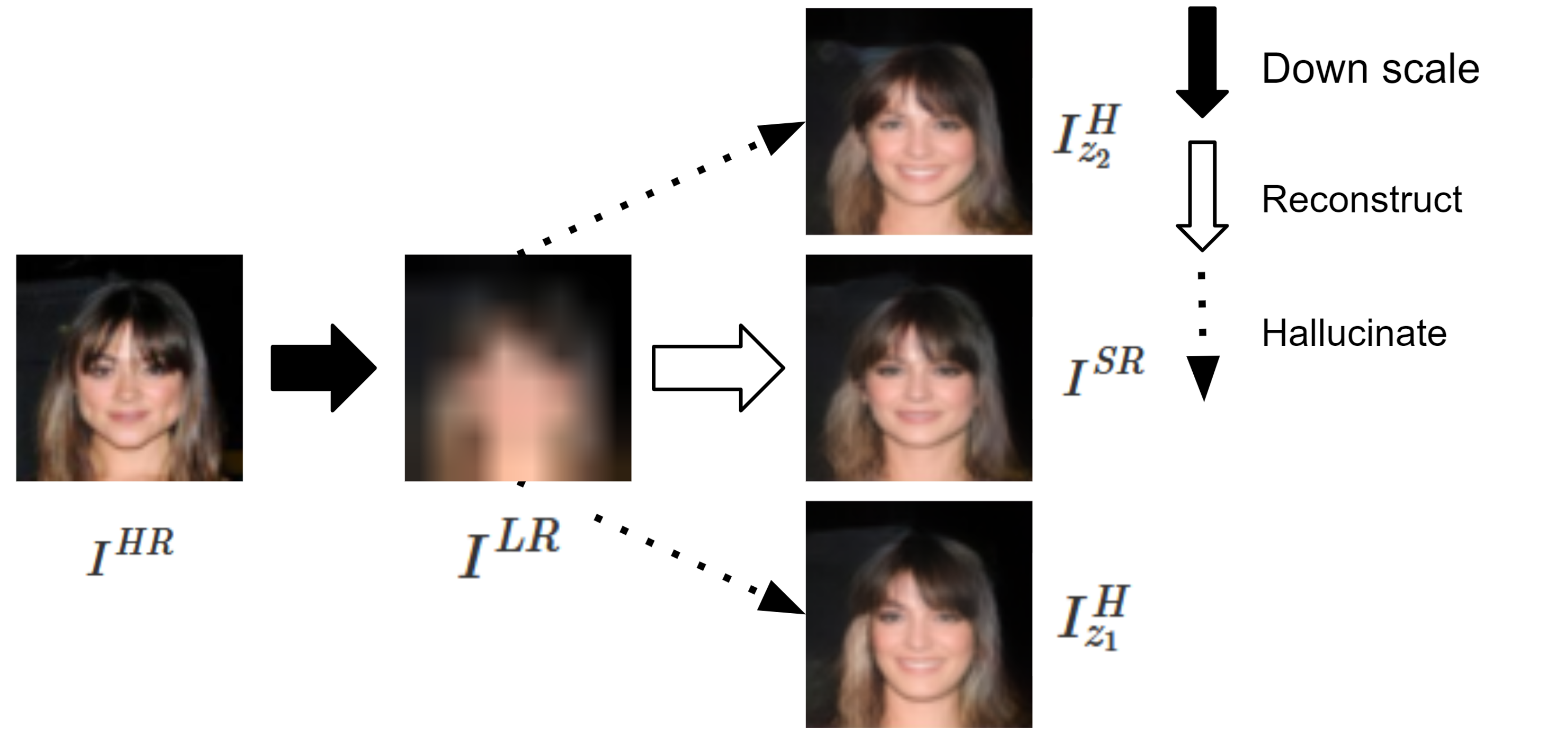}
    \caption{Conceptual representation of proposed approach: starting from an original image ($I^\mathit{HR}$) and its downscaled version ($I^\mathit{LR}$), different HR images can be generated, that is here the reconstructed image ($I^\mathit{SR}$) and two hallucinated ones ($I^\mathit{H}_{z_1}$ and $I^\mathit{H}_{z_2}$).}
    \label{fig:concept}
\end{figure}

Methods such as \cite{prevgrad} exploited the image’s gradient as structure guidance for super-resolution. This helps the model to better learn and preserve structural and textural information in the image. We follow a similar path but we are not limiting its use to preserve the structure of the image but to anchor the exploration over the image-gradient manifold in order to generate other images that would preserve most of the structure given by the image’s gradient. The main contribution of the paper lies in the novel neural architecture that makes use of high-frequency within the image not only to recover one HR image but also to generate a wide variety of plausible HR images that downscale to the same LR version of of this image -- see Fig.~\ref{fig:concept} that illustrates the concept of our approach.

\section{Approach}
\label{sec:format}

Given a LR image $I^\mathit{LR}$, the objective of the proposed approach is twofold:
\begin{itemize} 
    \item Reconstruct the ground truth HR image that allows the retrieval of its corresponding image gradient;
    \item Hallucinate a variety of HR images that match, when downscaled, the input LR image ($I^\mathit{LR})$. 
\end{itemize}
To tackle the aforementioned objectives, we propose a generator $G$ that takes as input pairs of $(I^\mathit{LR},\,z)$, where  $z$ is a $m$-dimensional white noise vector, $z \sim \mathcal{N}^m(0,1)$. 
We follow the same assumptions made in \cite{lag}, with our proposed model deemed to reconstruct the ground-truth HR image when $z=0$, otherwise ($z \neq 0$) sampling a variety of images matching the LR input but having a varied image gradient:
\begin{align}
G(I^\mathit{LR},\,0) & = (I^\mathit{SR},\,g^\mathit{SR})  \hspace{2.25em}\text{(reconstruction)},\label{eq:1}\\
G(I^\mathit{LR},\,z) & = (I^\mathit{H}_{z},\,g^{H}_{z})  \hspace{3em}\text{(hallucination)}.\label{eq:2}
\end{align}
In these equations, $I^\mathit{SR}$ is the reconstructed ground-truth image while $I^\mathit{H}_{z}$ is a hallucinated version given a noise vector $z$. Also, $g^\mathit{SR}$ and $g^\mathit{H}_{z}$ are respectively the reconstructed image gradient of the ground truth image and the hallucinated image gradient given a noise vector $z$.

\subsection{Image Reconstruction}

Like traditional SR models, the proposed model aims to minimize the distance between the ground-truth and the reconstructed HR images. In addition, our model aims to preserve structural information in the original image by using the ground-truth image $I^\mathit{HR}$ and its correspond gradient $g^\mathit{HR}$, computed as follows:
\begin{align}
g_{x} &=I^\mathit{HR}(x+1,\,y)-I^\mathit{HR}(x-1,\,y),\\
g_{y} &=I^\mathit{HR}(x,\,y+1)-I^\mathit{HR}(x,\,y-1),\\
g^\mathit{HR} &= \|(g_{x},\,g_{y})\|_2.
\end{align}
As shown in \eqref{eq:1}, the generator takes as input the pair $(I^\mathit{LR},\,0)$ and outputs $(I^\mathit{SR},\,g^{SR})$. The Perceptual loss \cite{perceploss} with VGG-16 \cite{Vgg} is then used by comparing the obtained $I^\mathit{SR}$ to the real $ I^\mathit{HR}$, by minimizing the Euclidean distances between the features of these images: 
\begin{equation}
\mathcal{L}_\mathit{percp}=\mathbb{E}_{I^\mathit{SR}}\left\|\phi_{i}(I^\mathit{SR})-\phi_{i}(I^\mathit{HR})\right\|_{1},\label{per}
\end{equation}
where $\phi_i$ is value output by the $i$-th layer of VGG.
As for comparing the gradient images, the mean absolute error of the images (pixels) is used:
\begin{equation}
\mathcal{L}_{grad}=\left\|g^\mathit{SR} - g^\mathit{HR}\right\|_{1}.\label{gradrecon}
\end{equation}
Finally, for both domains, we use an adversarial loss for assessing the quality of the images and gradients generated, that is a non-saturating logistic loss \cite{goodfellow} with $R_1$ regularization \cite{r1}.

\subsection{Image Hallucination}
Unlike the reconstruction, image hallucination does not rely on comparing the generated image to the ground truth image. Now in addition to the low-resolution input image $I^\mathit{LR}$, the proposed method uses a random vector $z$ to generate and explore other plausible solutions.  
However, it is worth noting that conditional GANs (cGANs) suffer from mode collapse problems \cite{sal,mode}. Also cGANs tend to ignore the random vector $z$ when conditioned on inputs that contain significant information about the output, for example in image-to-image translation models \cite{pix2pix, cyclegan} where the model tends to ignore the random vector. Without addressing this problem in cGANs, the hallucinated images would be equal to the reconstructed image and the model will regress to the same behaviour of one-to-one mapping. To circumvent this problem, we impose on the model a diversity constraint first introduced in \cite{diversity}, which is applied here only on the gradient output, to orient exploration of the structural and texture space of the gradient image.

To enforce the effect of $z$ for obtaining diverse solutions, the generator aims at \emph{maximizing} loss $\mathcal{L}_{z}(G)$: 
\begin{equation}
\mathcal{L}_{z}(G)=\mathbb{E}_{z_{1},z_{2} \sim \mathcal{N}^n(0,1)}\left[\min \left(\frac{d(g^{H}_{z_1},\,g^{H}_{z_2})}{\|z_{1}-z_{2}\|}, \tau\right)\right],
\end{equation}
where $d(\cdot,\cdot)$ is a distance metric, and $g^\mathit{H}_{z_1}$ and $g^\mathit{H}_{z_2}$ are the gradient images obtained from HR hallucinated images with $z_1$ and $z_2$ sampled from $\mathcal{N}^n(0,1)$ , respectively, both hallucinated images obtained from the same LR input. The minimum with $\tau$ is conducted to ensure numerical stability of the optimization, in case where the output is not bounded by an activation function (e.g., tanh or sigmoid). 

Maximizing $\mathcal{L}_{z}(G)$ encourages the generator to explore more the HR space, by producing a varied set of hallucinated samples. To make sure that the hallucinated samples stay faithful to the given LR input image, we impose the following constraint:
\begin{equation}
\|\mathrm{DS}(I^{H}_{z}) - I^{LR}\| < \epsilon,\label{eq:epsi}
\end{equation}
where $\mathrm{DS}(\cdot)$ is a down-scaling operation, and $\epsilon$ is a hyper-parameter.

\subsection{Overall Objective}

Two discriminators are used in the generative model, one for assessing the HR images ($D_{I}$) and the other to validate its corresponding gradient ($D_{g}$). For both networks, non-saturating logistic loss \cite{goodfellow} with $R_1$ regularization \cite{r1} is used:
\begin{align*}
\mathcal{L}_{adv}^{g} &= -\mathbb{E}_{g^\mathit{HR}}\left[\log(1-D_{g}(g^\mathit{HR}))\right] - \mathbb{E}_{g^\mathit{SR}}\left[\log D_{g}(g^\mathit{SR})\right],\\
\mathcal{L}_{adv}^{I} &= -\mathbb{E}_{I^\mathit{HR}}\left[\log(1-D_{I}(I^\mathit{HR}))\right] - \mathbb{E}_{I^\mathit{SR}}\left[\log D_{I}(I^{SR})\right].  
\end{align*}


As for generating reconstructed HR images, we use a weighted sum over $\mathcal{L}_\mathit{percp}$ \eqref{per}, $\mathcal{L}_\mathit{grad}$ \eqref{gradrecon} and the adversarial loss:
\begin{equation}
\mathcal{L}_\textit{recons} = \gamma(\mathcal{L}_\mathit{percp} + \mathcal{L}_\mathit{grad})
     + \beta (\mathcal{L}_{adv}^{g} + \mathcal{L}_{adv}^{I}).
\end{equation}
And for generating hallucinated HR images, the loss is:
\begin{equation}
    \mathcal{L}_\textit{halluc} = \mathcal{L}_{adv}^{g} + \mathcal{L}_{adv}^{I} + \alpha \mathcal{L}_{z}.
\end{equation}

\section{Implementation}

\begin{figure}[t]
    \centering
    \includegraphics[width=\linewidth]{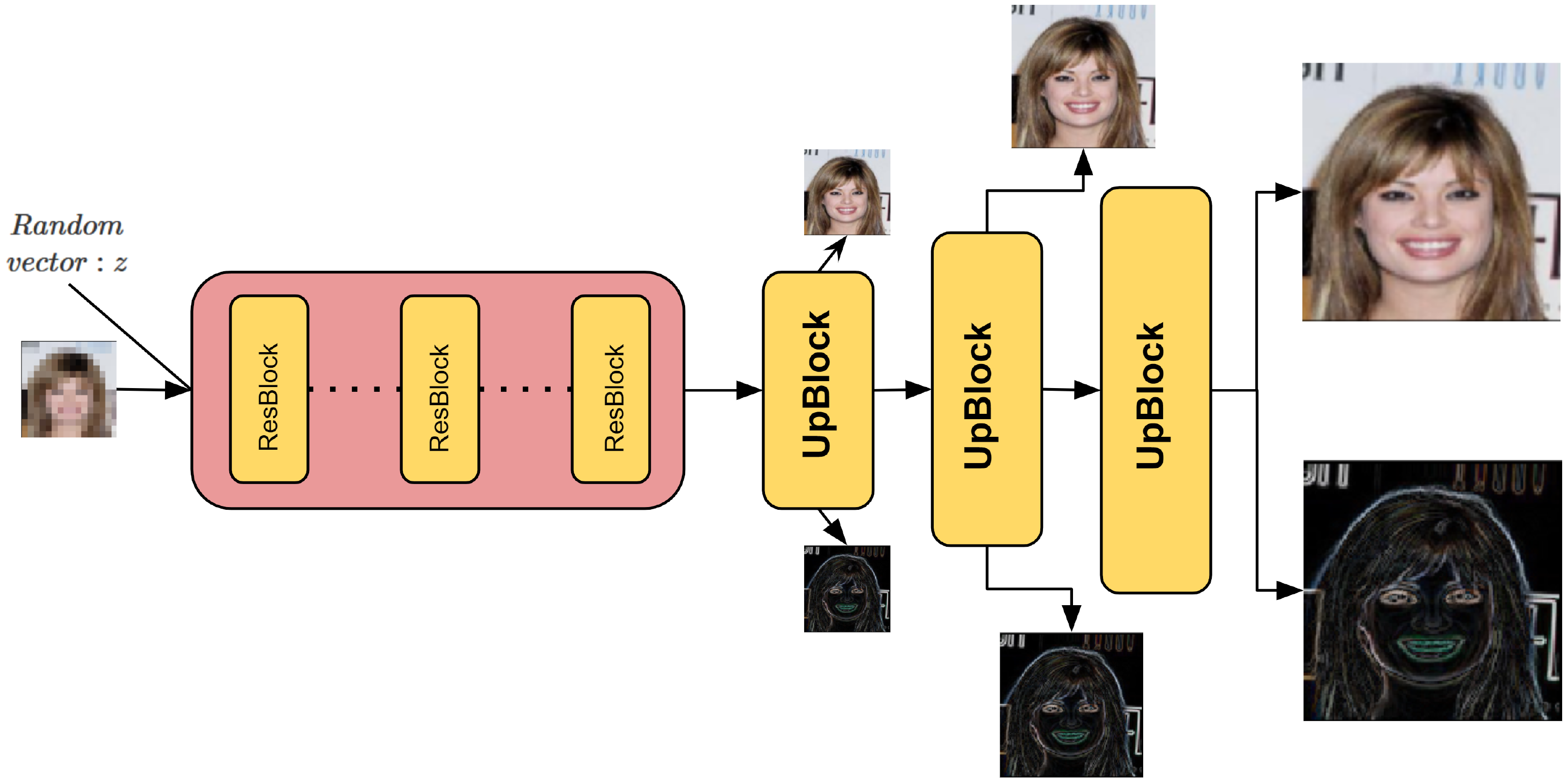}
    \caption{Generator Architecture: the network consists of 8 residual blocks followed by stacked upsampling blocks (Upblocks) -- see Fig.~\ref{fig:upblock} for the inner layers of UpBlock.}
    \label{fig:gen}
\end{figure}

\subsection{Conditioning}

We choose to condition the discriminator $D_I$ on the constraint \eqref{eq:epsi} by concatenating it at the last block of the network.
We follow \cite{lag} by rounding the downscaled fake image to the closest colour value and dividing it by $r = 2/255$ (since images are in the range $[-1,1]$).
\begin{equation}
F = \max\left( \frac{\left|\lfloor DS(I^\mathit{H}_z) \rceil - I^\mathit{LR}\right|}{r} - \epsilon,\,0\right).\label{eq:f}
\end{equation}
That prevents producing exceedingly large values for the discriminator’s weights in order to measure infinitesimal differences.
A straight through estimator is used to pass the backpropagation gradient through the rounding operation.
Therefore, the discriminator receives two inputs, fake/real images and $F$ given in \eqref{eq:f}:
$$
\left\{\begin{array}{ll}
D_I\left(I^\mathit{HR},\,0\right) & \text{for real images} \\
D_I\left(I^\mathit{H}_{z},\,F\right) & \text{for fake images} 
\end{array}\right..
$$

\subsection{Architecture}

Our model consists of a generator, that takes a LR image, a noise vector, and outputs two images: a HR image and its corresponding gradients. Two discriminators are also used during the learning, which are taking real/fake images and their gradients, respectively.

\begin{figure}[t]
    \centering
    \includegraphics[width=\linewidth]{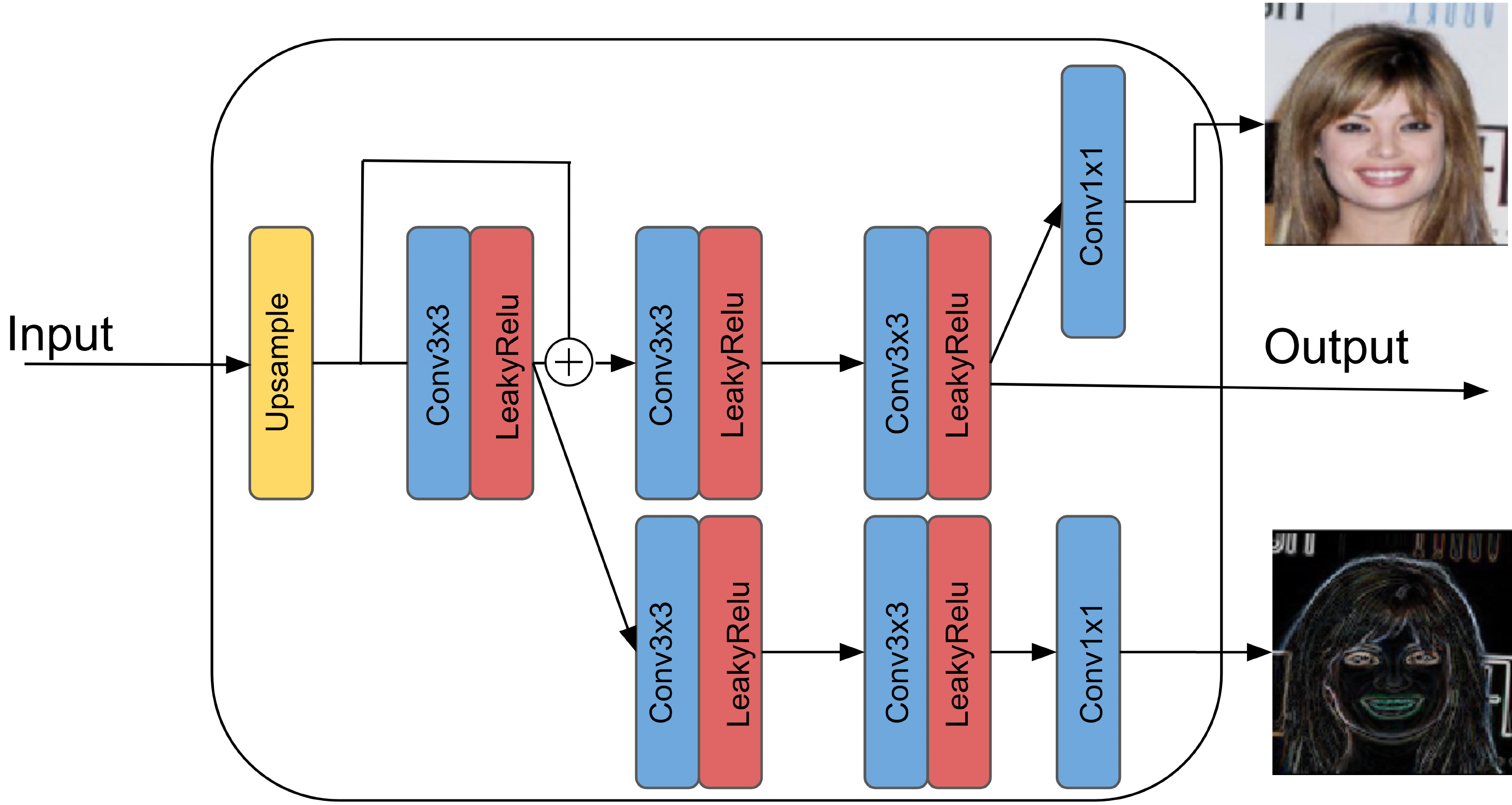}
    \caption{Upblock architecture: a convolution block shared between the two domains is used to enable the learning of image structures, which are then followed by two domain-specific convolution and a $1\times 1$ convolution to make a projection of the values in the feature space into the pixel space. The key in using a skip connection is to keep the low frequency information such as colours intact.}
    \label{fig:upblock}
\end{figure}

The generator global architecture is inspired by \cite{MSGGAN} (see Fig.~\ref{fig:gen}). First the residual blocks take the LR image with noise vectors, then they are followed by up-sampling blocks, where each block outputs an image and its corresponding gradient at that scale (see Fig.~\ref{fig:upblock}). This way the gradient flows to the generator at different scales, allowing it to be more stable, leading to a faster learning. 
As mentioned before, two discriminators are used, one for each domain ($I^\mathit{HR}$ and $g^\mathit{HR}$), their architecture is mirrored version of the generator, but since each discriminator is domain-specific, the upblock is replaced by a downblock, where we use first a $1\times 1$ convolution to pass the image from the pixel space to feature space. This is followed by another two convolutions, by leaky RELUs and then by average pooling for downsampling. After the downsampling blocks, the residual blocks are just like the generator. The discriminator directly outputs the features of last convolutions, without using linear layers.




\begin{figure*}[t]
\centering
\includegraphics[width=0.78\linewidth]{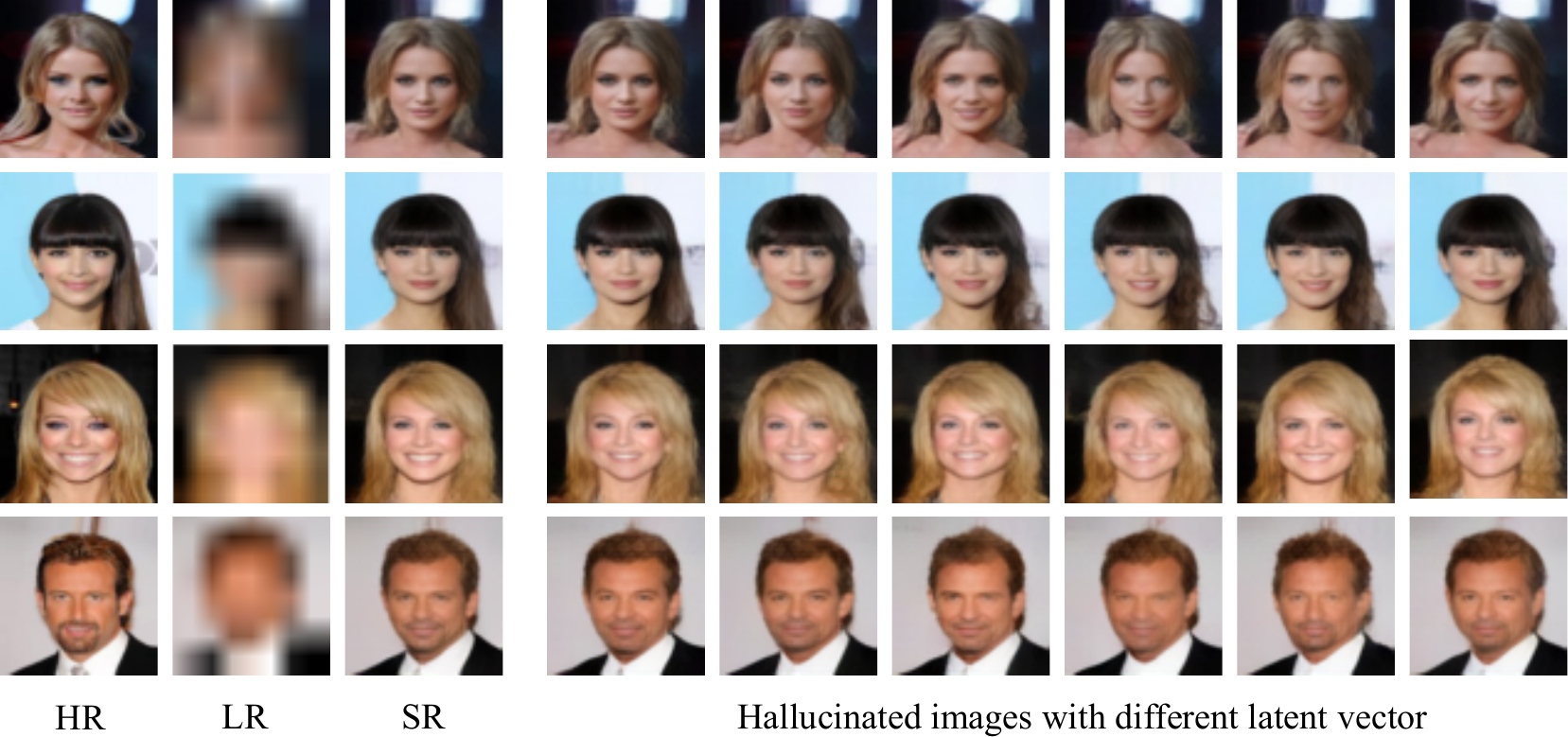}
\caption{Results on 8x scaling: reconstruction (SR) and hallucination using different $z$ vectors}
\label{fig:viz}
\end{figure*}

\section{Experiments}

\subsection{Settings}
Experiments were conducted on the CelebA \cite{celeba} dataset, to assess the capability of our model, which was trained on upscaling images 8 times and then compared to other SR methods (i.e, \cite{srgan} and \cite{esrgan}). 

Our model is trained with the Adam optimizer\cite{adam} ($lr = 1e^{-4}$, $beta_1= 0.$, $beta_2= 0.9$) with TTUR \cite{ttur}, using batch sizes equal to 8. We use a leaky RELU for all activation functions in the generator and both discriminators, with a leak equal to 0.2.

We fix $\epsilon$ in \eqref{eq:f} to $0.1$, while for $\mathcal{L}_\mathit{recons}$ we set $\gamma = 10$ and $\beta=0.1$, and for $\mathcal{L}_\mathit{halluc}$, $\alpha $ is equal to $1$.

\begin{figure}[t!]
    \centering
    \includegraphics[width=0.95\linewidth]{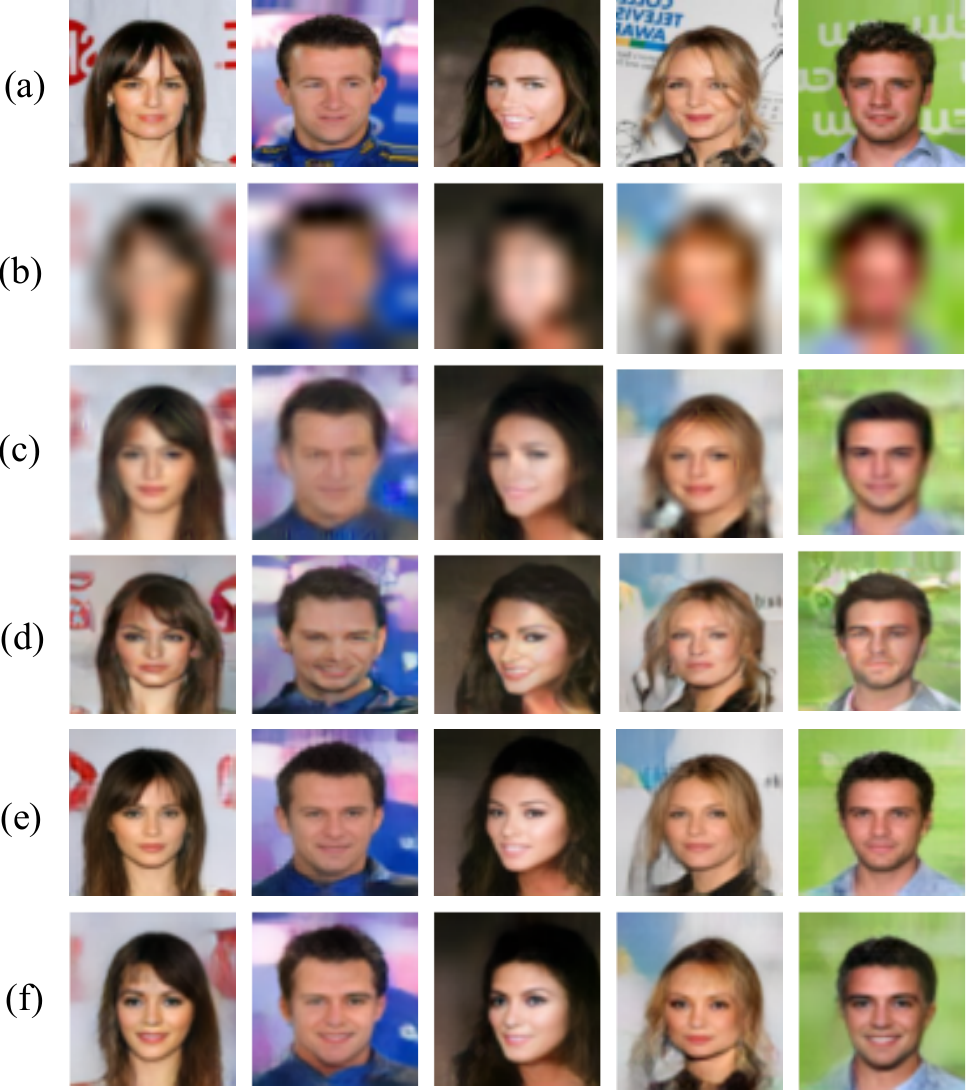}
    \caption{Results on CelebA 8x scaling: (a) ground truth, (b) bicubic reconstruction, (c) SRGAN, (d) Esrgan, (e) Ours : Reconstruction ($z=0$), and (f) Ours : Hallucination ($z \neq 0$).}
    \label{fig:im}
\end{figure}
\begin{table}[tb!]
\centering
\begin{tabular}{@{}lccc@{}}
\toprule
Method  & SSIM $\uparrow$ & PSNR $\uparrow$ & LPIPS $\downarrow$\\ \midrule
Bicubic & 0.43 & 18.11 & 0.51  \\
SRGAN \cite{srgan}   & 0.62 & 20.15 & 0.24  \\
ESRGAN \cite{esrgan}  & 0.55 & 18.43 & 0.23  \\
Ours    & \textbf{0.68} & \textbf{21.08} &\textbf{0.18}  \\ \bottomrule
\end{tabular}
\caption{Comparison of results for 8x scaling on CelebA \cite{celeba}, all results are reported on the test set.}
\label{tab:results}
\end{table}

\subsection{Results}
\label{sec:pagestyle}

We use the traditional SR metrics peak signal-to-noise ratio (PSNR), structural similarity index (SSIM) and learned perceptual image patch similarity (LPIPS) \cite{nossim}. Although our method is optimized to generate images with high perceptual quality, which means that the results do not correlate very well with PSNR and SSIM \cite {nossim}, we still achieve excellent performance, with an improvement by a large margin on all metrics -- see Table~\ref{tab:results} for details.


We report the results of our model and compare these results with the results obtained using the approaches described in \cite{srgan,esrgan}.
We show that when the input is a very low resolution image, where most of the information is missing, the use of the gradient has a huge impact on the results.
As illustrated in Fig.~\ref{fig:im}, the reconstructions of our model are closer to the real images than other methods, but also more coherent since the structural properties of the images is being preserved by making the model learn the gradient of the image.

In addition to having better reconstruction given the use of $\epsilon = 0.1$ for \eqref{eq:epsi} in our experiments, the model has more freedom to hallucinate other semantics that do not appear in the low-resolution image, such as changing the facial expressions or even changing the person's identity (see Fig.\ \ref{fig:viz}).


\section{Conclusion}
In this paper we propose a new super-resolution method that makes use of novel GAN  architecture that is able to recover images from extremely low resolution images, and hallucinate a wide variety of other possibilities.  We showed the superiority of our approach in both tasks on the CelebA dataset.

\newpage
\bibliographystyle{IEEEbib}
\bibliography{strings,refs}

\end{document}